\documentclass{article}

\usepackage{arxiv}

\usepackage[utf8]{inputenc} 
\usepackage[T1]{fontenc}    
\usepackage{hyperref}       
\usepackage{url}            
\usepackage{booktabs}       
\usepackage{amsfonts}       
\usepackage{nicefrac}       
\usepackage{microtype}      
\usepackage{lipsum}
\usepackage{tabularx}
\usepackage{bm}
\usepackage{graphicx}
\usepackage{amsmath}
\usepackage{placeins}
\usepackage{lscape}

\title{Spatiotemporal mapping of malaria prevalence in Madagascar using routine surveillance and health survey data}

\begin{document}
\maketitle\vspace{-0.25cm}

\begin{center}
\fontsize{10pt}{16pt}\fontseries{b}\selectfont Rohan Arambepola$^{1*}$,   Suzanne H. Keddie$^{2}$,   Emma L. Collins$^{1}$,   Katherine A. Twohig$^{1}$, Punam Amratia$^{1}$, Amelia Bertozzi-Villa$^{1,8}$, Elisabeth G. Chestnutt$^{1}$, Joseph Harris$^{2}$, Justin Millar$^{1}$, Jennifer Rozier$^{2}$,   Susan F. Rumisha$^{1}$, Tasmin L. Symons$^{1}$, Camilo Vargas-Ruiz$^{1}$,   Mauricette Andriamananjara$^{5,7}$,   Saraha Rabeherisoa$^{5}$, Ars\`ene C. Ratsimbasoa$^{5,6}$, Rosalind E. Howes$^{1,4}$, Daniel J. Weiss$^{1,2,3}$, Peter W. Gething$^{1,2,3}$ and Ewan~Cameron$^{1,2,3}$\par
\end{center}

\small$^{1}$Big Data Institute, Li Ka Shing Centre for Health Information and Discovery, University of Oxford, Oxford, United Kingdom; $^{2}$Telethon Kids Institute, Perth Children's Hospital, Perth, Australia; 
$^{3}$Curtin University, Perth, Australia; 
$^{4}$Foundation for Innovative New Diagnostics, Geneva, Switzerland; 
$^5$Programme National de Lutte contre le Paludisme, Antananarivo, Madagascar; 
$^6$University of Fianarantsoa, Fianarantsoa, Madagascar; 
$^7$Minist\`ere de Sant\'e Publique, Antananarivo, Madagascar; 
$^8$Institute for Disease Modeling, Bellevue, WA, USA\vspace{0.5em}
\\$^*$\textbf{For correspondence}: rohan.arambepola@stx.ox.ac.uk

\noindent\rule{\textwidth}{0.5pt}\vspace{0.3cm}

\normalsize

\fontsize{10pt}{11pt}\selectfont

\begin{abstract}
Malaria transmission in Madagascar is highly heterogeneous, exhibiting spatial, seasonal and long-term trends. Previous efforts to map malaria risk in Madagascar used prevalence data from Malaria Indicator Surveys. These cross-sectional surveys, conducted during the high transmission season most recently in 2013 and 2016, provide nationally representative prevalence data but cover relatively short time frames. Conversely, monthly case data are collected at health facilities but suffer from biases, including incomplete reporting.

We combined survey and case data to make monthly maps of prevalence between 2013 and 2016. Health facility catchments were estimated and incidence surfaces, environmental and socioeconomic covariates, and survey data informed a Bayesian prevalence model. Prevalence estimates were consistently high in the coastal regions and low in the highlands. Prevalence was lowest in 2014 and peaked in 2015, highlighting the importance of estimates between survey years.  Seasonality was widely observed. Similar multi-metric approaches may be applicable across sub-Saharan Africa.
\end{abstract}


\section{Introduction}
Malaria is a major public health problem in Madagascar, with an estimated 2.16~million cases leading to more than 5000~deaths in the country in 2018~\cite{world2019world}. Malaria burden decreased in the early 2000s with an increase in control efforts but this progress was largely halted following political turmoil in 2009 \cite{barmania2015madagascar, bhatt2015effect, howes2016contemporary} resulting in a resurgence in endemicity in the last decade~\cite{howes2016contemporary, kang2018spatio, world2019world, ihantamalala2018spatial}.

Malaria transmission is highly heterogeneous across Madagascar, reflecting the diverse ecological landscape of the island. Transmission is highest in the east and west coastal regions, where it follows a seasonal pattern with clinical incidence peaking between February and May depending on location. In the central highlands and the desert south, transmission is lower and annual trends are less consistent, with temporal variation appearing to be driven by outbreak dynamics~\cite{howes2016contemporary, randrianasolo2010sentinel, nguyen2020mapping}. Howes et al.~\cite{howes2016contemporary} identified eight contiguous ecozones representing distinct transmission settings (Central highlands, Highlands fringe west, Highlands fringe east, Northeast, Northwest, Southeast, Southwest, South) which are shown in Figure~\ref{pfpr}(a). When devising the 2018--2022 malaria National Strategic Plan, the National Malaria Control Programme (NMCP) of Madagascar classified 106 of the 114 districts in Madagascar as control areas, 3 pre-elimination and- 5 elimination, based on reported case numbers in 2016~\cite{nsp18}. Control strategies are currently stratified by risk level, with intermittent preventative treatment for pregnant women (IPTp) and mass distribution of insecticide-treated bed nets (ITNs) in the 106 control districts and indoor residual spraying (IRS) targeted at high transmission districts in the southeast and southwest. In pre-elimination settings the focus is on outbreak and active case detection~\cite{president2019mdg, president2018mdg}.

Routine malaria case data are collected through the Health Management Information System (HMIS) from reports from health facilities. These data are collected monthly and have a high spatial coverage (see Figure~\ref{API}) but also suffer from a number of potential biases. This passive case detection is unlikely to capture all malaria cases in the community, missing those that do not seek care or do so from informal or private providers, which likely represents a large fraction of the population (treatment-seeking rates in the public sector in 2013 were estimated to be around 45\%)~\cite{battle2016treatment, mis2013, mis2016}. Furthermore, cases seen at health facilities may not be diagnosed or reported to the central system due to resource constraints (such as malaria rapid diagnostic test (RDT) stock-outs) or weak communication infrastructure~\cite{howes2016contemporary}. Nevertheless, while these data may not represent all malaria incidence, they are an important source of information on trends in transmission due to their high temporal and spatial coverage~\cite{howes2016contemporary, ihantamalala2018spatial, nguyen2020mapping, bennett2014methodological, chanda2012impact}.

Malaria Indicator Surveys (MIS) provide another source of data for understanding the spatiotemporal patterns of malaria endemicity~\cite{mis2011, mis2013, mis2016}. These cross-sectional surveys are designed to be nationally representative and collect data on a number of indicators, including prevalence of  \textit{Plasmodium falciparum} infection in individuals between 6 and 59 months of age. They are conducted with a standardised methodology applied over all sites and surveys and, unlike routine case data, are not affected by reporting incompleteness, treatment-seeking behaviour or standards for clinical diagnosis. For these reasons, prevalence information from national health surveys has traditionally been the primary source of data for mapping malaria risk in sub-Saharan Africa~\cite{battle2019mapping, weiss2019mapping, bhatt2015effect, kang2018spatio}. Kang et al.~\cite{kang2018spatio}~used a Bayesian hierarchical model to map prevalence in Madagascar in 2011, 2013 and 2016 using parasite prevalence data from the 2011, 2013 and 2016 MIS reports~\cite{mis2011, mis2013, mis2016}. However, cross-sectional surveys provide limited insight into seasonal patterns of transmission or transmission in years when no survey took place. Variation in the timing of surveys between years may also make it difficult to distinguish changes in prevalence between survey years from seasonal variation. Moreover, survey data is less informative in low burden areas, where sample sizes are likely to be inadequate to accurately assess changes in transmission due to low rates of detectable parasitaemia~\cite{sturrock2016mapping}.

In this study, we combined routine case data and prevalence survey data within a formal modelling framework, taking advantage of their relative strengths, in order to provide a more complete understanding of the spatiotemporal heterogeneity of transmission between 2013 and 2016. From the routine case data, we produced spatially smooth monthly incidence surfaces which were then used as a covariate in a Bayesian geostatistical model of prevalence. This approach is similar to the use of modelled prevalence  surfaces to map incidence~\cite{lucas2019model} and the use of other modelled surfaces, such as temperature suitability~\cite{weiss2014air} or accessibility~\cite{weiss2018global}, as inputs when mapping malaria risk~\cite{battle2019mapping, weiss2019mapping, bhatt2015effect, kang2018spatio}. The use of both incidence and prevalence data is also similar to the work of Lucas et al.~\cite{lucas2020mapping} combining prevalence and aggregated incidence data in a joint model. However, the two-step modelling process used here is conceptually simpler than a joint model and allows the relationship between incidence and prevalence to be learned within the prevalence model (rather than imposing a previously estimated relationship, such as \cite{cameron2015defining}). An important benefit of learning this relationship is that systematic biases in routine case data are implicitly accounted for. Modelling the incidence and prevalence processes separately is equivalent to making a `cut' between the incidence and prevalence processes in the model. Modularising the inference in this way also prevents over-reliance on the less reliable incidence data in the final prevalence estimates~\cite{liu2009modularization, jacob2017better}. We incorporated routine case data to update the estimates of prevalence for 2013 and 2016 made by Kang et al.~\cite{kang2018spatio} and to produce estimates for 2014 and 2015. By producing monthly prevalence maps over all four years, both long term and seasonal trends in prevalence across the country could be assessed. We used the eight ecozones identified by Howes et al.~\cite{howes2016contemporary} as a basis for assessing how these trends vary spatially.

\section{Methods}
\subsection{Study data}
\begin{figure}
\includegraphics[width=0.95\linewidth]{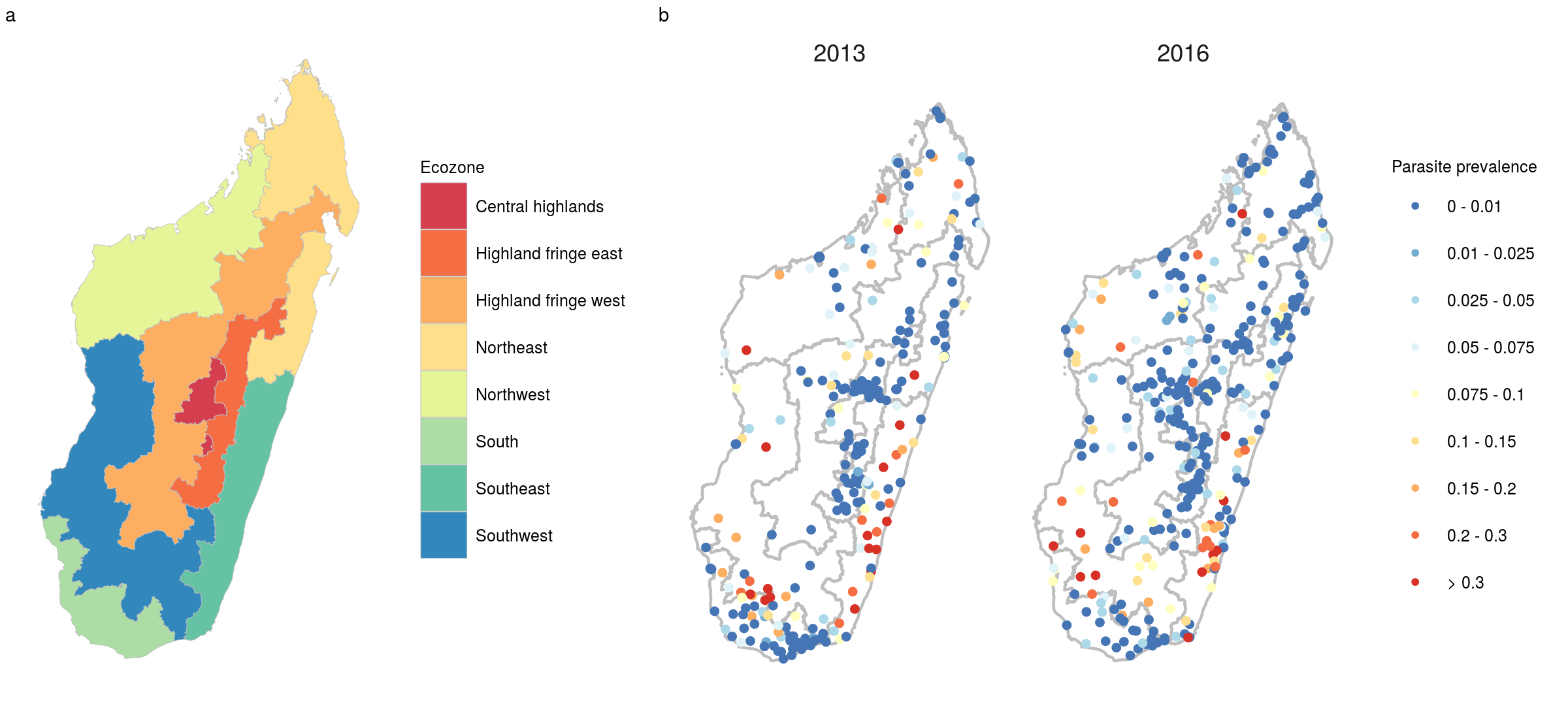}
\caption{(a) Ecozones defined by Howes et al.~\cite{howes2016contemporary} representing contiguous areas with distinct patterns of transmission. (b) Prevalence rates at survey sites in the 2013 and 2016 MIS.}
\label{pfpr}
\end{figure}

Parasite prevalence data was available from Malaria Indicator Surveys that took place in Madagascar in 2013 and 2016. These data consist of geo-located clusters where the prevalence of \textit{P. falciparum} infection was measured, determined by microscopy, in individuals between 6 and 59 months of age ($\textit{Pf}\textrm{PR}_{6-59\textrm{mo}}$). These surveys were conducted with a standard protocol and survey sites were selected to produce nationally representative estimates of prevalence in this age group. The number of positive individuals and the total number tested was recorded at each site. In the 2013 and 2016 surveys, 6323 and 6927 individuals were screened across 274 and 358 sites, respectively. The survey sites and proportions of positive individuals are shown in Figure~\ref{pfpr}(b) and full details can be found in the original reports~\cite{mis2013, mis2016}.

Monthly health facility case data between January 2013 and December 2016 were provided by the NMCP. These data come from HMIS data reports and represent clinical cases of malaria confirmed by an RDT for all ages, irrespective of parasite species. Data were available from 3342 health facilities in Madagascar, of which 2801 were geo-located and verified using a separate dataset of geo-located health facilities from the Institut Pasteur de Madagascar (as described in \cite{nguyen2020mapping}).

A number of covariates were used to inform the prevalence model, which are detailed in Table~\ref{cov_table}. Most of these variables were environmental, influencing vector abundance, parasite behaviour and environmental suitability. Two of these covariates (accessibility to cities~\cite{weiss2018global} and night lights~\cite{elvidge2017viirs}) relate to the development and urbanicity of a location and therefore are related to vector abundance and to socioeconomic factors, such as access to healthcare. In total there were 26 potential features for the prevalence model -- 8 static covariates and 6 dynamic covariates, each considered at 0, 1 and 2 month time lags. Causal feature selection~\cite{arambepola2020nonparametric} was used to select the variables included in the final model.

\begin{table}
\fontsize{10pt}{14pt}\selectfont
\caption{\label{cov_table}List of covariates. }
    \begin{tabularx}{\columnwidth}{p{7em}Xlp{5.5em}p{5em}}
\hline
Covariate & Description  & Type & In any causal feature set & In final feature set\\ \hline
Rainfall	~\cite{funk2014quasi}		&	Climate Hazards Group Infrared Precipitation with Station Data	&	Dynamic	&	Lag 0	&	Lag 0	\\
LST day	~\cite{modisLST}		&	Daytime land surface temperature	&	Dynamic	&	No	&	No	\\
LST night	~\cite{modisLST}		&	Night-time land surface temperature	&	Dynamic	&	Lag 1	&	No	\\
TCB 	~\cite{modisTCB}		&	Tasselled cap brightness; measure of land reflectance	&	Dynamic	&	Lag 2	&	No	\\
EVI	~\cite{modisEVI}		&	Enhanced vegetation index	&	Dynamic	&	No	&	No	\\
TSI Pf	~\cite{weiss2014air}		&	Temperature suitability index for \textit{P. falciparum}	&	Dynamic	&	Lag 2	&	No	\\
Accessibility	~\cite{weiss2018global}		&	Distance to cities with population $>$ 50,000	&	Static	&	Yes	&	Yes	\\
AI	~\cite{trabucco2009global}		&	Aridity index	&	Static	&	Yes &	Yes	\\
Elevation	~\cite{farr2007shuttle}		&	Elevation as measured by the Shuttle Radar Topography Mission (SRTM)	&	Static	&	Yes	&	No	\\
PET	~\cite{trabucco2009global}		&	Potential evapotranspiration	&	Static	&	Yes	&	No	\\
Slope	~\cite{farr2007shuttle}		&	GIS-derived surface calculated from SRTM elevation surface	&	Static	&	Yes	&	No	\\
Night lights	~\cite{elvidge2017viirs}		&	Index that measures the presence of lights from towns, cities and other sites with persistent lighting	&	Static	&	No	&	No	\\
Distance to water	~\cite{lehner2004global}		&	GIS-derived surface that measures distance to permanent and semi-permanent water based on presence of lakes, wetlands, rivers and streams, and accounting for slope and precipitation	&	Static	&	Yes	&	Yes	\\
TWI	~\cite{farr2007shuttle}		&	Topographic wetness index	&	Static	&	No	&	No

    \end{tabularx}
    \label{table: simulation parameters}
\end{table}

\subsection{Catchment population}
In order to calculate incidence rates at each health facility, estimates for catchment populations (the number of people likely to seek treatment at each facility) were needed.
A catchment model was used to estimate these (treatment-seeking) catchment populations, based on travel time to health facilities. The country was divided into a grid of approximately 5km-by-5km pixels. For each pixel, the travel time to each health facility was calculated using a friction surface (defining travel time through each pixel) developed by Weiss et al.~\cite{weiss2018global} and a least cost algorithm \cite{dijkstra1959note}. Given that an individual in pixel $i$ seeks treatment, the probability they seek treatment in health facility $j$, $p(\mathrm{pixel}_i\rightarrow\mathrm{HF}_j)$, was modelled as inversely proportional to square of the travel time to that health facility. That is,
$$p(\mathrm{pixel}_i\rightarrow\mathrm{HF}_j) = \frac{t(\mathrm{pixel}_i\rightarrow\mathrm{HF}_j)^{-2}}{\sum_{j=1}^{N_\mathrm{HF}}t(\mathrm{pixel}_i\rightarrow\mathrm{HF}_j)^{-2}}$$
where $t(\mathrm{pixel}_i\rightarrow\mathrm{HF}_j)$ represents the travel time from pixel $i$ to health facility $j$ and $N_\mathrm{HF}$ was the total number of health facilities. The catchment population of health facility $j$ was then calculated as
$$\sum_{i=1}^{N_\mathrm{pixel}} \mathrm{population}_i \times p(\mathrm{pixel}_i\rightarrow\mathrm{HF}_j)$$
where $\mathrm{population}_i$ is the treatment seeking-adjusted population of pixel $i$ and $N_\mathrm{pixel}$ is the total number of pixels. The proportion of the population in each pixel who would seek treatment at any health facility was also modelled as a function of travel time, which has been identified as an important factor in treatment-seeking behaviour for fever in Madagascar \cite{pach2016qualitative}. This proportion was modelled as a logistic function (similar to the functional forms considered by Alegana et al.~\cite{alegana2012spatial})
$$\frac{\alpha}{1 + \exp(\sigma t)} + \beta$$
where $t$ is the travel time to the nearest health facility in minutes. The parameters values $\alpha=0.6$, $\sigma=0.00916$, $\beta=0.15$ were chosen such that the maximum and minimum possible treatment-seeking proportions were 0.6 and 0.15, and the treatment-seeking proportion at $t=120$ minutes was 0.3. These parameter values were chosen to produce a similar relationship between treatment-seeking and travel time as that observed in the 2013 and 2016 MISs~\cite{mis2013, mis2016} (see supplementary material Figure 3) and match overall estimated treatment-seeking rates~\cite{battle2016treatment}. Catchment populations were calculated for each year between 2013 and 2016 using hybrid population surfaces from the Gridded Population of the World v4 \cite{GPWv4} and WorldPop \cite{tatem2017worldpop}. In order to assess the sensitivity of the final estimates to these modelling assumptions, three alternative sets of populations were considered: catchment populations supplied by the NMCP (based on the 1993 national census adjusted by a fixed annual growth rate \cite{moh2014} with no adjustment for treatment-seeking behaviour) and two sets of catchment populations generated by the catchment model under different treatment-seeking parameters (see supplementary material). The analysis was repeated using these catchment populations and the resulting prevalence estimates were compared.

\subsection{Incidence model}
Routine case data from health facilities were modelled using a Bayesian geospatial model to produce monthly incidence surfaces which were then used as inputs to the prevalence model. Let $c_{it}$ be the number of cases observed in month $t$ ($t=$1, ..., 48) at health facility~$i$~($i=$1, ..., $N$), which is at location $s_i$ and has a treatment-seeking catchment population $E_i$. The number of cases observed was modelled as a Poisson process
$$c_{it}\sim \mathrm{Pois}(E_i\times\lambda_{it})$$
with mean equal to the product of the catchment population and the underlying incidence rate, $\lambda_{it}$. The log incidence rate was modelled as
$$\log \lambda_{it} = \beta_0 + f(s_i, t)$$
where $\beta_0$ was an intercept and for each month $f(\cdot, t)$ was a realisation of a Gaussian process over space with zero mean and M\'atern covariance structure. The M\'atern covariance function is parameterised by the range, $\rho$, and  marginal variance, $\sigma^2$, the values of which were chosen by a search over parameter space to maximise accuracy when predicting incidence in held out locations. These parameters were optimised jointly over all months. The value of smoothness parameter, $\nu$, was fixed at 1.

\subsection{Prevalence model}
Prevalence data from MIS surveys were also modelled using a Bayesian geospatial model, informed by environmental and socioeconomic covariates and the modelled incidence surfaces. Let $y_i$ be the number of infected individuals and $N_i$ be the number of individuals tested in survey $i$ ($i=1$, ..., $M$), taking place in location $s_i$ at time $t_i$. The results of the survey were modelled as a realisation of a binomial process
$$y_i \sim \mathrm{Binomial}(p_i, N_i)$$
with underlying prevalence $p_i$ at this location and point in time. The logit-transformed prevalence was modelled as 
$$\mathrm{logit}(p_i) = \beta_0 + \bm{\beta}^TX_i + \beta_0^{\mathrm{inc}}g(\lambda(s_i, t_i)) + \beta_1^{\mathrm{inc}}g(\lambda(s_i, t_i + 1)) + f(s_i)$$
where $\beta_0, \bm{\beta}, \beta_0^{\mathrm{inc}}, \beta_1^{\mathrm{inc}}$ were parameters (with $\beta_0^{\mathrm{inc}}$ and $\beta_1^{\mathrm{inc}}$ non-negative), $X_i$ were covariate values and $\lambda(s, t)$ was the $\log$ incidence value from the modelled incidence surfaces at location $s$ and time $t$. $f$ was modelled as a realisation of a Gaussian process over space with M\'atern covariance, parameterised by the range $\rho$ and marginal variance $\sigma^2$, while $g$ is a realisation of a Gaussian process over incidence with a squared exponential kernel, parameterised by the scale $\kappa$ (with fixed variance 1).  This allowed a non-linear effect of incidence on logit-prevalence while limiting model-complexity by assuming a priori that this relationship is smooth (by the choice of squared exponential kernel and placing an appropriate prior on the scale parameter). This flexibility was important as incidence is likely to be the main driver of prevalence in the model (while the additional covariates are likely to be less informative and therefore were modelled as having linear effects) and assuming a linear effect produces prevalence-incidence relationships that do not match empirical observations \cite{cameron2015defining}. Incidence in the subsequent month was included in addition to incidence in the current month, as the presence of parasites in the blood could result in a clinical case that is recorded in a health centre several weeks later. The parameters $\beta_0^{\mathrm{inc}}$ and $\beta_1^{\mathrm{inc}}$ allowed the model to learn the relative predictive power of incidence at these two time points.

The Bayesian model was completed by placing appropriate priors on the parameters and hyperparameters. Normal priors were placed on $\beta_0, \bm{\beta}, \beta_0^{\mathrm{inc}}, \beta_1^{\mathrm{inc}}$ centred around 0 with standard deviation 1 for $\beta_0$ and 0.25 for $\bm{\beta}, \beta_0^{\mathrm{inc}}, \beta_1^{\mathrm{inc}}$. The M\'atern covariance parameters were given log-normal priors with mean 3 and standard deviation 0.1 for $\rho$ and mean 0 and standard deviation 0.1 for $\sigma$, shrinking the spatial term towards a fairly flat and smooth field. A log-normal prior was placed on $\kappa$ with mean 3 and standard deviation 0.1. Both the incidence and prevalence models were fitted using the Gaussian Markov Random Field approximation \cite{lindgren2011explicit} with the \textsf{Template Model Builder} package~\cite{kristensen2015template} in \textsf{R} and samples were drawn from a Laplace approximation to the posterior to produce associated uncertainty estimates. The model was validated by fitting the model with data from the 2013 survey and making predictions for 2016 (to evaluate model performance when predicting prevalence in previously unobserved time points) and through $k$-fold cross validation.

\subsection{Feature selection}
The covariates included in the prevalence model were selected from the potential covariates described in Table~\ref{cov_table} using causal feature selection~\cite{arambepola2020nonparametric, guyon2007causal}. In total there were 29 potential features, 18 dynamic (6 covariates each at 0, 1 and 2 month time lags) and 8 static. The idea behind causal feature selection is to select features with the most direct causal relationships to the response based on the available data. 

We describe the procedure briefly here which is described in full detail by Arambepola et al. \cite{arambepola2020nonparametric}. The PC algorithm~\cite{spirtes2000causation}, a causal discovery algorithm, was used to infer causal relationships between the different features and malaria prevalence. In particular, the output of this algorithm identified all direct causes of prevalence. To quantify the certainty of these direct causes, the algorithm was repeatedly applied to bootstrapped samples of the data. The certainty of a feature being a direct cause of prevalence was then quantified as the proportion of repeats in which it was inferred to be a direct cause. For a given minimum certainty, feature sets were then generated by selecting all direct causes of prevalence and a number of potential feature sets were generated by varying the minimum certainty required between 0 and 1. Out of these potential feature sets, the final set chosen was the feature set which maximised the cross-validated predictive performance of the model. The causal discovery algorithm relies on conditional independence testing. We used the Randomized Conditional Independence Test \cite{strobl2019approximate} to perform scalable non-parametric conditional independence tests.

Selecting causal features may be beneficial for a number of reasons. It is possible that causal selection may lead to smaller feature sets, especially in situations in which many features are associated with the response variable but relatively few are directly causal~\cite{guyon2007causal}. Small feature sets may improve computational efficiency and reduce overfitting. Models built on causal feature sets may also be more
robust to common problems such as concept drift and covariate shift \cite{scholkopf2012causal} and therefore make more useful predictions further forward in time or in previously unobserved locations. Arambepola et al.~\cite{arambepola2020nonparametric} showed that using causal feature selection resulted in improved performance when forecasting malaria incidence compared to classical feature selection.

\section{Results}
\subsection{Catchment populations} 
The median size of the modelled catchment populations was 2890 (LQ: 1710, UQ: 4740). The total population served according to the modelled populations increased from 10.15 million in 2013 to 11.02 million in 2016, corresponding to approximately 43\% of the Malagasy population each year, which is largely in agreement with estimated treatment-seeking rates \cite{battle2016treatment}.

\subsection{Incidence}
\begin{figure}
\includegraphics[width=0.95\linewidth]{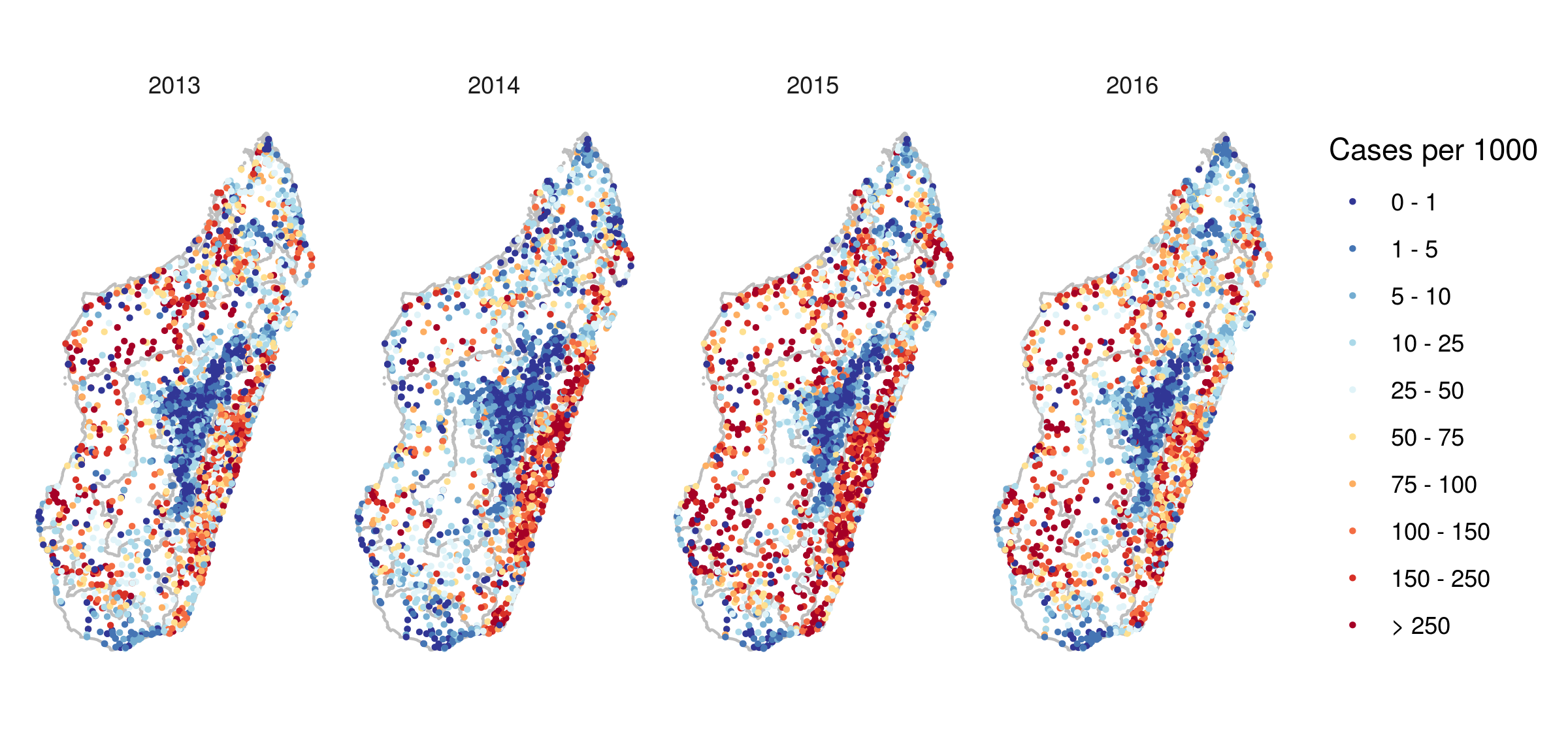}
\caption{Annual incidence rates at each health facility based on routine case data and modelled catchment populations.}
\label{API}
\end{figure}

\begin{figure}
\includegraphics[width=0.95\linewidth]{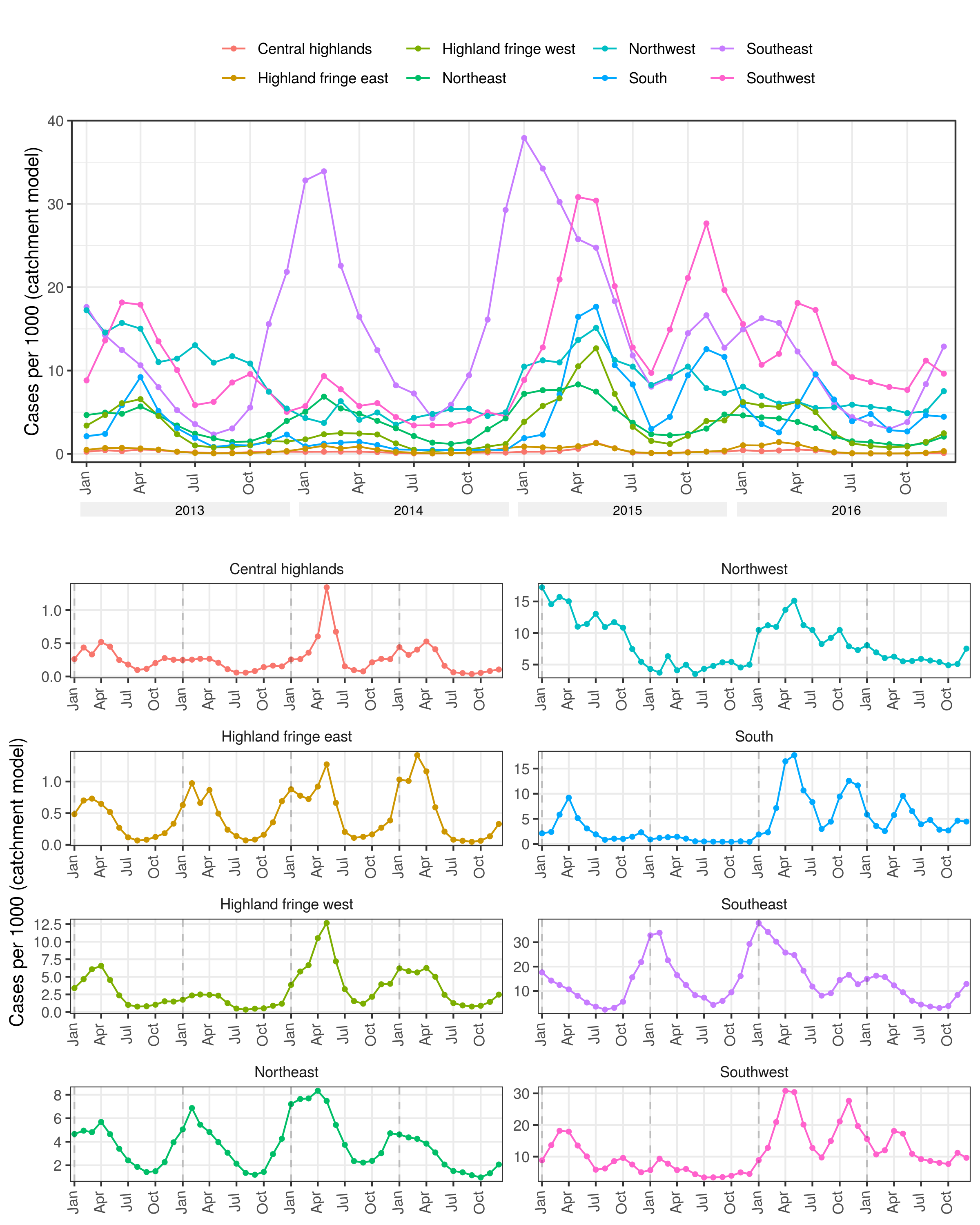}
\caption{Monthly incidence rates at each health facility based on routine case data and modelled catchment populations stratified by ecozone (2013-2016).}
\label{inc_month}
\end{figure}

Annual incidence rates at each health facility (calculated using reported cases and modelled catchment populations) are shown in Figure \ref{API}. Spatial patterns of incidence in 2013 and 2016 were similar to observed prevalence (Figure \ref{pfpr}(b)) with lower rates in the central highlands and the south, and higher rates in the east and west coastal regions. Compared to 2013, incidence in 2016 was generally higher in health facilities in the southeast and southwest but lower in the north. Across the four years, rates were generally lowest in 2014 (with the exception of the east coast) and highest in 2015. Figure \ref{inc_month} shows monthly incidence rates aggregated by ecozone. A seasonal pattern of a single peak in incidence between January and April can be seen to some extent in most regions, most clearly in the Southeast, Northeast and (despite low overall incidence) Highland fringe east ecozones. An increase in cases in 2015 was observed in almost all ecozones.

\begin{figure}
\includegraphics[width=0.95\linewidth]{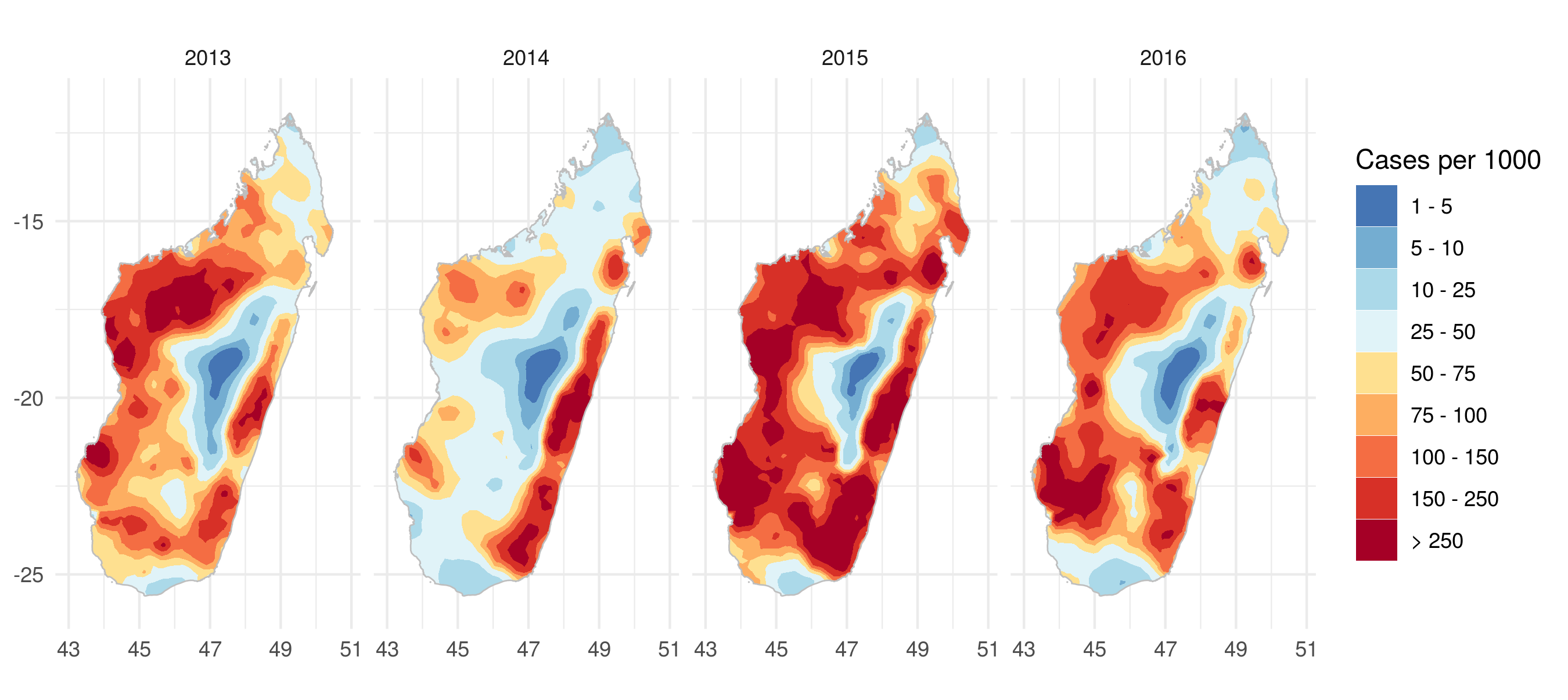}
\caption{Incidence surfaces for all ages aggregated annually.}
\label{API_smooth}
\end{figure}

The optimal hyperparameters for the incidence model were $\rho=e^{-0.1}$ and $\sigma=e^{-2}$. Figure \ref{API_smooth} shows the smooth incidence surfaces produced by this model aggregated annually, which as expected reflect the overall spatiotemporal trend of the observed data. It should be noted that these surfaces were only intended to be a spatial smoothing of the treatment seeking-adjusted case data reflecting relative spatial trends, rather than an enumeration of true incidence rates, and are an intermediate step in the modelling process rather than an output. 

\subsection{Prevalence}
The causal discovery algorithm identified 4 dynamic and 4 static feature sets which combined to give 16 potential feature sets for the prevalence model. The features that were present in at least one of these sets are listed in Table \ref{cov_table} and these sets are listed in full in the supplementary material. The features used in the final model were rainfall~\cite{funk2014quasi} with no time lag, accessibility to cities~\cite{weiss2018global}, an aridity index~\cite{trabucco2009global} and distance to water~\cite{lehner2004global}. The posterior mean and 95\% credible interval for all the model parameters, including coefficients of these features, are shown in Table \ref{param_table}. As expected, incidence was the most important predictor, with incidence in the current month having a greater effect than incidence in the following month. The other covariates in the model had smaller and less consistent effects on prevalence. When fitting the model on 2013 data and making predictions for 2016, there was a correlation of 0.63 between predicted and observed values. For 3, 4 and 5-fold cross-validation, there were average correlations of 0.58, 0.59 and 0.6, respectively. These values are reasonably high given that the observed prevalence rates are themselves noisy point estimates of underlying prevalence; for context, the standard errors of the mean of binomial samples of 20 individuals (the median survey size) with true prevalence values of 0.01, 0.05 and 0.15 are 0.022, 0.049 and 0.080, respectively.

\begin{table}[!h]
\fontsize{10pt}{14pt}\selectfont
\centering
\caption{\label{param_table}Mean and 95\% credible intervals for the prevalence model parameters.}
\begin{tabular}{lcll} \hline
Parameter & Mean & CI\\ \hline
Intercept&	-3.956	&	-6.082,	-1.926	\\
Accessibility&	0.060	&	-0.021,	0.150	\\
AI&	-0.066	&	-0.172,	0.042	\\
Distance to water&	-0.031	&	-0.108,	0.042	\\
Rainfall (no lag)&	0.019	&	-0.097,	0.128	\\
$\log\rho$&	1.48	&	1.263,	1.693	\\
$\log\sigma$&	-0.762	&	-0.968,	-0.561	\\
$\log\lambda_\mathrm{inc}$&	1.883	&	0.881,	3.043	\\
$\beta_0^\mathrm{inc}$&0.802	&	0.407,	1.22	\\
$\beta_1^\mathrm{inc}$&0.497	&	0.184,	0.879	
\end{tabular}
\end{table}

\begin{figure}
\includegraphics[width=0.95\linewidth]{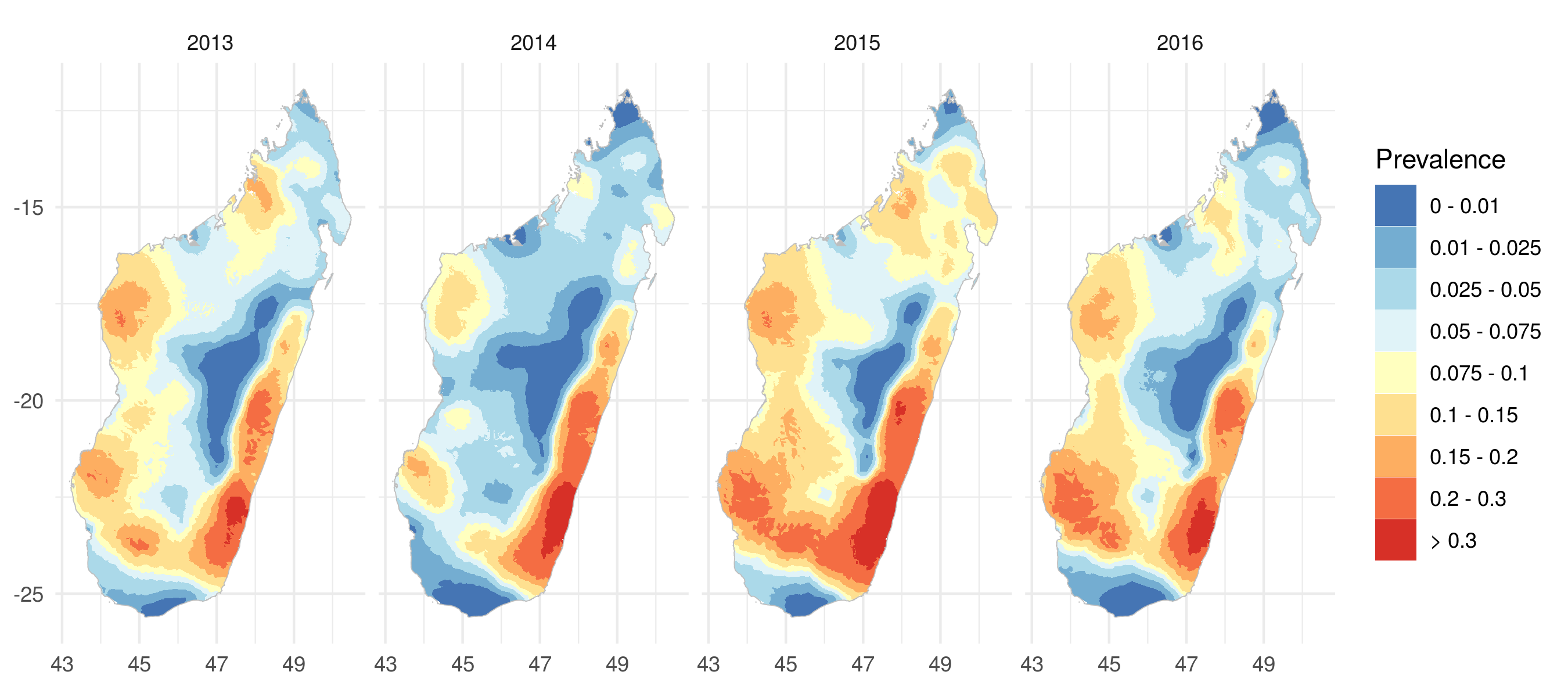}
\caption{Annual prevalence estimates for individuals between 6 and 59 months of age.}
\label{prev_annual}
\end{figure}

\begin{figure}
    \centering
    \includegraphics[width=0.8\linewidth]{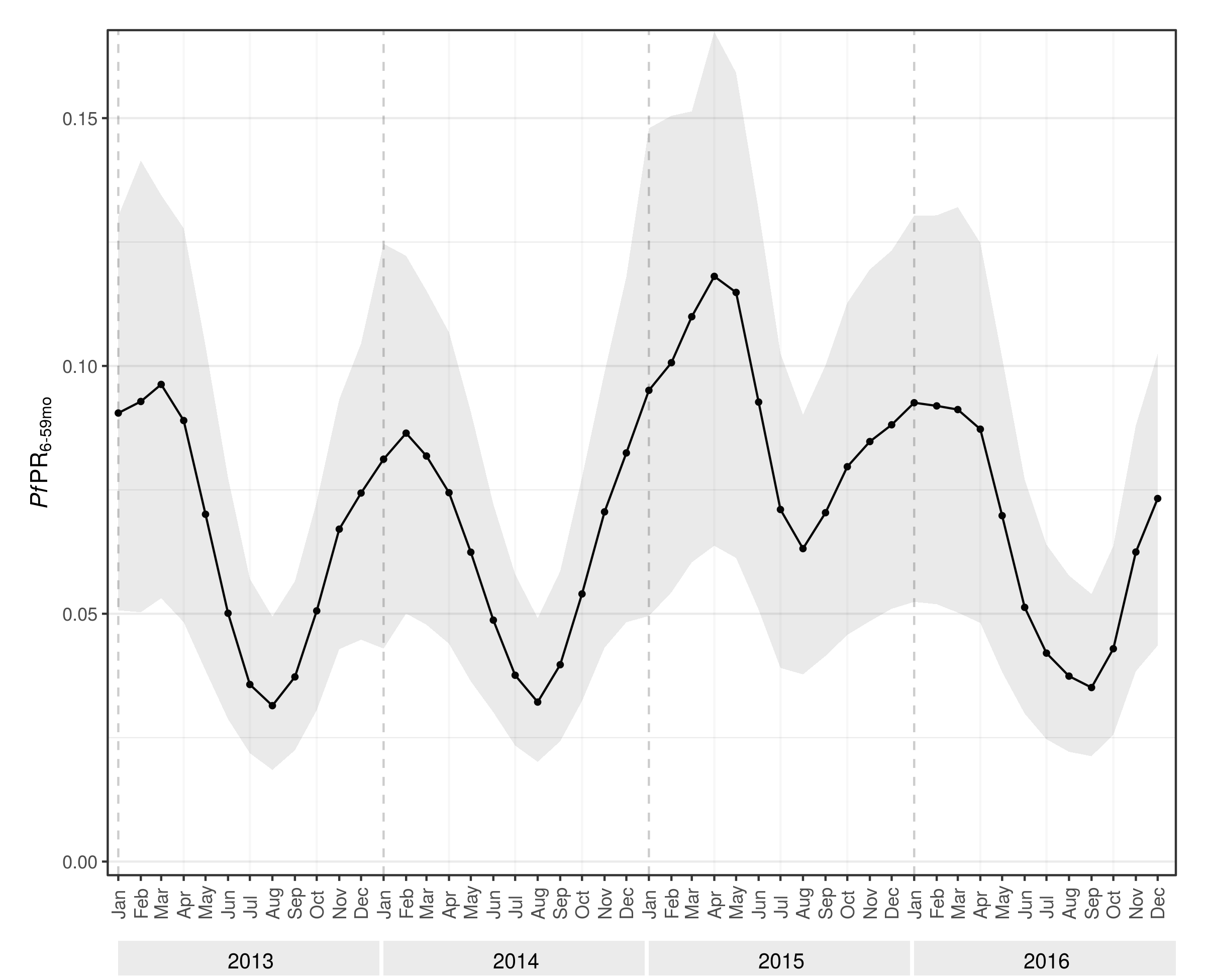}
    \caption{Population-weighted mean parasite prevalence over time with 95\% credible interval.}
    \label{prev_ts}
\end{figure}

Prevalence estimates for individuals between 6 and 59 months of age are shown aggregated annually in Figure \ref{prev_annual}. Spatial patterns were similar across all four years, with highest estimated prevalence near the southeast coast and other areas of high prevalence along the west coast. Prevalence estimates were consistently low in the far south of the country and in the central highlands. Prevalence in the north varied more between years, although was generally low or moderately low. Population-weighted mean prevalence was lowest in 2014 (6.0\%, 95\% credible interval (CI): 3.3-8.8), followed by 2013 (6.4\%, CI: 3.5-9.5) and 2016 (6.6\%, CI: 3.6-9.6), and was highest in 2015 (8.9\%, CI: 4.6-12.9). Population-weighted prevalence over time (Figure \ref{prev_ts}) showed a clear seasonal pattern, peaking between February and April each year with the lower prevalence occurring between July and September. Prevalence was highest in 2015, with the peak prevalence across all four years occurring in April 2015 (11.4\%, CI: 5.8-16.0) and notably high prevalence sustained into the lower transmission season. By the second half of 2016, prevalence appeared to have returned to similar levels to 2013 and 2014.

\begin{figure}
    \centering
    \includegraphics[width=0.95\linewidth]{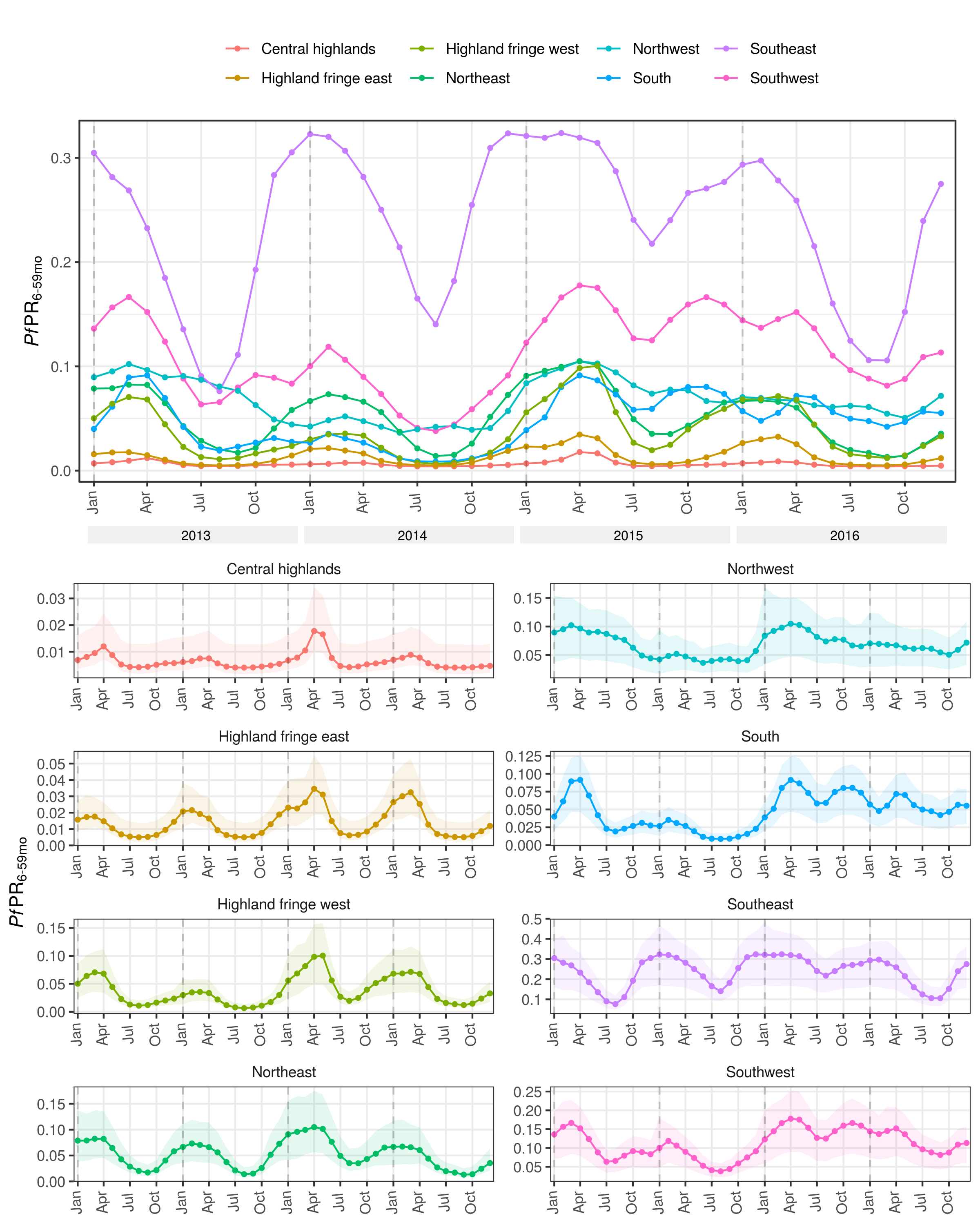}
    \caption{Population-weighted mean prevalence over time stratified by ecozone with 95\% credible interval.}
    \label{prev_ts_ecozone}
\end{figure}

\begin{figure}
    \centering
    \includegraphics[width=0.7\linewidth]{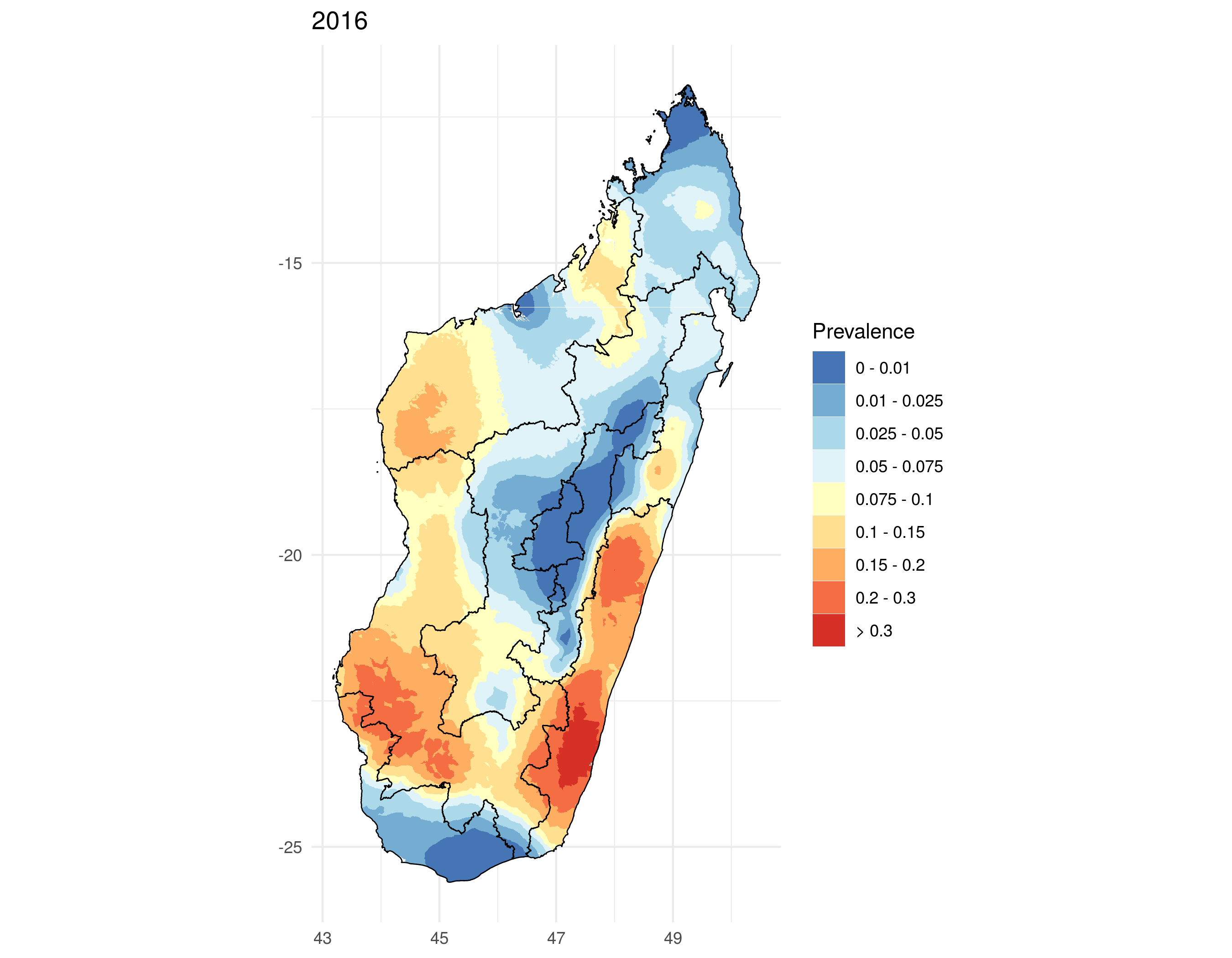}
    \caption{Prevalence estimates for 2016 with ecozones borders shown in black.}
    \label{2016_ecozone}
\end{figure}

Figure \ref{prev_ts_ecozone} shows the mean prevalence over time for each ecozone. Prevalence was consistently highest in the Southeast ecozone, with peak monthly prevalence of 0.3 or more in all years, while consistently lowest in the Central highlands and Highland fringe east ecozones. Seasonal patterns of prevalence are present in most of these ecozones, including both higher transmission regions (Northeast, Southeast and Southwest) and the lower transmission highland ecozones. These regional seasonal trends are similar to the pattern observed for mean prevalence across the country, with a single peak occurring between February and April in most areas (though often earlier in the Southeast ecozone) and lowest prevalence around August. The Northwest and South ecozones exhibit some similar seasonal trends but these are less consistent. In most areas, prevalence in 2014 was slightly lower than in 2013, with large decreases in the Northwest and South ecozones. Increased prevalence in 2015 was observed in all areas. In the highland ecozones, prevalence was higher than normal around April but later in the year returned to similar levels to previous years. In the rest of the country, a high peak in April was generally followed by higher than average prevalence throughout the rest of the year and into 2016. The average prevalence in 2016 (also shown in Figure \ref{prev_annual}) is shown again in Figure \ref{2016_ecozone} with ecozone borders highlighted. The ecozones generally correspond well to these latest prevalence estimates. The greatest heterogeneity within ecozones was in the Highland fringe west ecozone, which contained a region of higher prevalence to the south, and the Northwest ecozone. 

\begin{figure}
    \centering
    \includegraphics[width=\linewidth]{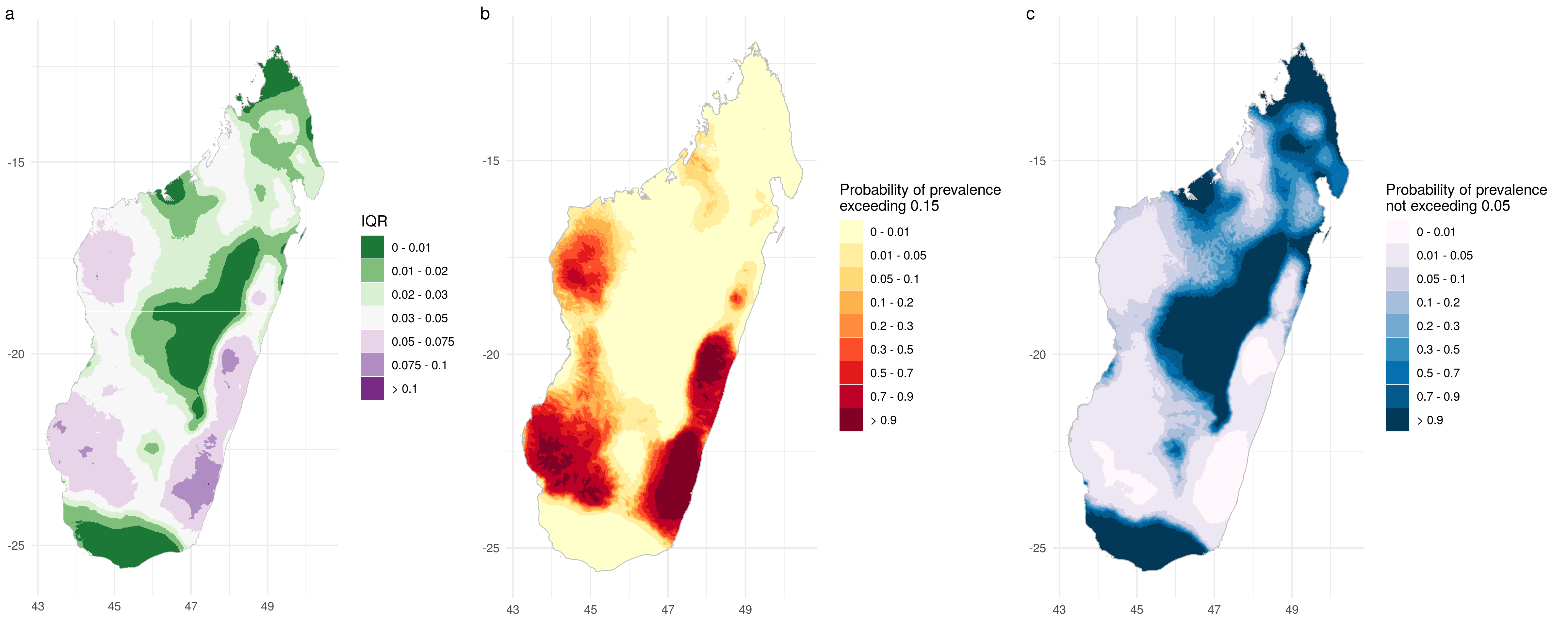}
    \caption{Uncertainty in 2016 annual prevalence estimates expressed with (a) interquartile range, (b) probability of prevalence exceeding 0.15 and (c) probability of prevalence not exceeding 0.05.}
    \label{2016_uncertainty}
\end{figure}

Figure \ref{2016_uncertainty} shows two ways of visualising the uncertainty in prevalence estimates, using annual 2016 estimates as an example. The first (Figure \ref{2016_uncertainty}(a)) maps the interquartile range of the posterior distribution. The spatial patterns of uncertainty are similar to the prevalence estimates, with higher (absolute) uncertainty areas of high prevalence. An alternative quantification of uncertainty is in terms of probability of estimates exceeding (or not exceeding) chosen values. Figure \ref{2016_uncertainty}(b) shows the probability of prevalence (sampled from the posterior distribution) exceeding 0.15. This can be interpreted as the probability that the true prevalence in 2016 exceeded 0.15. In the majority of the country there is a very high certainty that prevalence was below 0.15, while on the east and west coasts there is fairly high confidence that prevalence exceeded this value. In the southwest, in particular, the probability of exceeding 0.15 is often above 90\%. Similarly, we can visualise the probability of not exceeding a certain value, as shown in Figure \ref{2016_uncertainty}(c) with the probability of not exceeding 0.05. Here we can identify areas in the highlands, south and far north where there is very high confidence of low prevalence. We also see more areas around the north where there is less certainty, with probabilities between 20-50\% of prevalence not exceeding 0.05. Maps of uncertainty in 2013, 2014 and 2015 using the interquartile range and exceedance probabilities are included in the supplementary material (supplementary material Figures 5-7).

The sensitivity analysis performed using different catchment populations showed that our prevalence estimates were fairly robust to different methods of estimating these populations. Despite the systematic differences between these estimates, leading to significantly different incidence estimates (supplementary material Figure 1), the resulting prevalence estimates were very similar (supplementary material Figure 2).

\section{Discussion}
Our results highlight the benefits of combining routine case data and cross-sectional survey data to provide a more complete understanding of prevalence over time. While the 2013 and 2016 MIS data show similar levels of prevalence in these two years, the routine case data suggests there was substantial variation in transmission between these time points, with decreasing case numbers in 2014 followed by a marked increase in 2015. Incorporating this case data into a formal modelling framework allowed us to quantify these trends in terms of changes in prevalence over time. The robust two-step modelling approach allowed us to learn from these noisy data sources effectively while retaining flexibility in important model components, in particular the incidence-to-prevalence relationship. 

By producing monthly risk maps we are able to assess temporal trends of interest in detail. For example, in much of the country the increase in prevalence in 2015 was caused by increased prevalence throughout the year and into early 2016, rather than only higher than average prevalence during the high transmission season. The observed increase in cases in 2015 has been attributed to cyclones and the resulting flooding and lack of supplies in many parts of the country~\cite{president2017mdg}. Our results may help confirm this or identify other factors driving these increases in transmission (and similarly the decreases in 2014). We can also distinguish regions where prevalence appears to have returned to normal levels by the second half of 2016 from regions where increased transmission appears to have been sustained, such as the South and Highland fringe west ecozones. The latter may be potential targets for increased control efforts. Monthly prevalence estimates allowed us to evaluate seasonal trends in prevalence. These trends were largely similar across the country, with one seasonal peak occurring between February and April. This seasonality was clear in many of the high transmission areas but was also observed in highland ecozones, despite the low overall prevalence in these regions. A better understanding of baseline transmission patterns in low transmission settings may improve outbreak detection~\cite{randrianasolo2010sentinel} or influence plans for moving towards pre-elimination and elimination in these areas.

We can also consider the long term trends over the four years in the context of the ITN mass distribution campaigns (MDCs) which have taken place approximately every three years since 2009 \cite{president2018mdg}. The second MDC took place in November 2012 on the east coast and October 2013 in the rest of the country and the third from September to December 2015~\cite{president2013mdg, president2014mdg}. Although continuous ITN distribution was carried out to supplement these campaigns, a decline in net coverage and effectiveness over time could be a factor in the overall increase in estimated burden up to the end of 2015.

When comparing our results to the MIS data, the prevalence estimates from the MIS reports of 9.1\% in 2013 \cite{mis2013} and 7.0\% in 2016 \cite{mis2016} are higher than our annual point estimates (6.4\% and 6.6\% respectively) but are consistent with our credible intervals. However, our results show limited evidence of a decrease in prevalence in 2016 compared to 2013. It appears likely that the decrease observed in the raw data is a consequence of the timing of the 2016 survey which primarily took place between May and July, around a month later than the 2013 survey and slightly past the seasonal peak in prevalence for the year. The annual 2013 and 2016 risk maps made by Kang et al.~\cite{kang2018spatio} show a similar spatial pattern to our results, with low rates of prevalence in the highlands and south of the country and higher rates in the east and west coasts. However, Kang et al.~\cite{kang2018spatio} estimate high prevalence in the north of the country, whereas in our results areas of higher prevalence are largely in the northeast. Similar to our results, their estimates show very similar prevalence in 2013 and 2016, although in general the overall mean estimated is slightly higher (9.3\% in 2016) than our point estimates. We can also compare the monthly estimates of prevalence in each ecozone. While in many regions overall seasonal trends were largely similar, the monthly prevalence estimates made by Kang et al.~\cite{kang2018spatio} show less consistent seasonal patterns with greater uncertainty. We would expect the monthly estimates in our analysis to be more robust, particularly outside of the survey months (April - June in 2013 and May - July in 2016) and in lower transmission areas, as there is more data to inform each month.

Comparing prevalence estimates to the routine case data, we can see that the temporal patterns in prevalence are typically smoothed trends from the case data (due to the spatial smoothing when producing the incidence surfaces and the temporal smoothing from using incidence at two time points to the predict prevalence). However, we can see distinct spatial trends in the prevalence estimates, for example prevalence was consistently highest in the Southeast ecozone in the whole study period despite similar or higher observed incidence rates in the Southwest ecozone from April 2015 onwards.

As well as visualising uncertainty using the interquartile range, we have used maps of exceedance (and non-exceedance) probabilities. In practice, the latter may be more interpretable and therefore more useful for communicating uncertainty. By choosing relevant thresholds, exceedance surfaces may be useful for identifying areas that are estimated to be high transmission with high confidence or areas that are most likely to be considered pre-elimination.


\subsection{Limitations}
A key modelling input in this study was monthly incidence rates at each health facility, which depend on the estimated catchment populations. Our model for estimating these catchment populations is based only on travel time to each health facility and does not take into account other factors that may influence treatment-seeking or choice of facility, such as type of facility~\cite{do2018associations, pach2016qualitative}. However, the results of our analysis when using catchment population estimates from the NMCP and when varying treatment-seeking behaviour (see supplementary material Figures 1 and 2) demonstrate that our modelling approach is fairly robust to differences in catchment population estimates. This is likely due to the spatial smoothing of incidence data and learned incidence-to-prevalence relationship. Treatment-seeking behaviour may also vary throughout the year, which is not accounted for in this analysis and may bias monthly estimates of prevalence. Modelling these temporal dynamics is extremely challenging, with limited treatment-seeking data available and behaviour that is likely to vary on a small spatial scale based on local infrastructure and geography.

A strength of our approach is the ability to account for bias in reporting of case information due to the learned relationship between prevalence and incidence. However, differences in reporting across the country cannot be accounted for in this relationship (which is constant across time and space). Howes et al.~\cite{howes2016contemporary} investigated the spatial variation in reporting in 2014 by looking at RDT stock-outs and proportion of distributed RDTs for which any result was reported to the HMIS by district. The proportion of RDT results reported was generally higher on the west coast, and therefore prevalence in this region may be somewhat overestimated, but across the rest of the country there were no strong spatial trends in reporting. Similarly, increases in reported cases over time may be partly due to increased access to healthcare or availability of RDTs, leading to overestimates of prevalence in more recent years, although the use of survey data in the first and last years of our study period should mitigate this.

Although the case data used in this study was for a different age range (all ages) to the estimates of prevalence (6-59 months), we believe this is unlikely to bias our estimates. There was a strong relationship between the number of cases in individuals under 5 and all ages in the routine case data (correlation of 0.91 across all health facilities and months) and the relationship between incidence and prevalence was learned within the model. Therefore we would expect the predictive power of incidence in either age range to be similar. Incidence for all ages was chosen due to the larger sample sizes in the data, which should produce more reliable estimates of incidence.

The aim of this study was to estimate prevalence, however estimates of incidence may be of more use for public health purposes (for example, district-level risk stratification by the NMCP is based on estimated incidence~\cite{nsp18}). Future work could therefore focus on mapping incidence. A common approach for generating incidence estimates is to use prevalence estimates and an established prevalence-incidence relationship~\cite{weiss2019mapping, bhatt2015effect}. This approach is generally used where prevalence estimates are informed only by prevalence survey data (where routine case data is unavailable or too unreliable to be used) but a similar transformation could be applied to the prevalence estimates produced here. A more direct way of combining routine case data and survey data to estimate incidence would be to use a joint model~\cite{lucas2020mapping}. To do so successfully would likely require a more complete understanding of the spatiotemporal trends in treatment-seeking and reporting completeness in order to account for bias in the routine case data due to these factors.

\section{Conclusion}
In this study, we used routine case data and survey data to produce monthly estimates of malaria prevalence between 2013 and 2016. Our results suggest that while malaria endemicity was similar in 2013 and 2016, there was considerable variation in the intervening period, with a small decrease in 2014 followed by a substantial increase in 2015. In many areas, this increase in prevalence was sustained throughout 2015 and early 2016, and in the Northwest and South ecozones prevalence had yet to return to 2014 levels by the end of 2016. Seasonality of transmission (with a single peak around March) was widely observed in high transmission areas and in some low transmission areas. The relative spatial patterns of prevalence were mostly consistent over time, with highest prevalence in the southwest and high prevalence in the southeast. Prevalence was consistently low in the highlands and in the south of the country. 

The Bayesian modelling approach applied here allowed us to make prevalence estimates with associated measures of uncertainty by learning temporal trends in transmission from the routine case data and calibrating these trends to prevalence observations from the survey data. These risk estimates could be an important tool for assessing the impact of control measures and the progress made towards the goals set out by the NMCP \cite{president2019mdg, president2018mdg} and for better understanding the drivers of changes in transmission. Our results demonstrate the considerable additional information that can be gained by combining data sources in this way. While previous risk mapping efforts produced monthly prevalence maps for 2011, 2013 and 2016~\cite{kang2018spatio}, these were informed by a relatively small number of prevalence observations from surveys which only covered 3 months of each year. We were able to make monthly maps of prevalence over a four year period (including years in which no surveys took place), informed by a large amount of monthly case data in addition to the survey data. Our sensitivity analysis demonstrated that these estimates are robust to varying assumptions about treatment-seeking behaviour and reporting incompleteness. 

Incomplete reporting, treatment-seeking behaviour, and inconsistent standardisation of clinical diagnoses affect the quality of routine case data in many countries of sub-Saharan Africa~\cite{alegana2020routine}. Consequently, the World Health Organization estimates for malaria burden are often based on community prevalence surveys alone~\cite{world2019world, weiss2018global}. The methods presented here may be a useful starting point for incorporating routine case data into estimates of malaria burden more widely.

\section*{Acknowledgements}
The authors thank Fanjasoa Rakotomanana and her team at Institut Pasteur de Madagascar for sharing their database of health facility geo-location data for validation of the NMCP data as previously described by Nguyen et al.~\cite{nguyen2020mapping}.

\section*{Additional information}
\subsection*{Competing interests}
The authors declare that they have no competing interests.

\subsection*{Author's contributions}
RA, EC and PWG conceived the study. RA, ELC, SHK and KAT designed and carried out the analysis. ELC, SHK, MA, SR and ACR prepared the datasets. EC, PWG and REH advised on the analysis. RA wrote the manuscript. All authors contributed to the interpretation of results. All authors read and approved the final manuscript.

\subsection*{Funding}
The first author was supported in this work through an Engineering and Physical Sciences Research Council (EPSRC) (https://epsrc.ukri.org/) Systems Biology studentship award (EP/G03706X/1). 
 
Work by the Malaria Atlas Project on methods development for Malaria Eradication Metrics including this work is supported by a grant from the Bill and Melinda Gates Foundation (OPP1197730).

\subsection*{Availability of data and materials}
Prevalence datasets, sample case data, and code are available at \url{https://github.com/rarambepola/Prevalence-Madagascar}. The raw case data that support the findings of this study are available from the Programme National de Lutte Contre le Paludisme de Madagascar and the Institut Pasteur de Madagascar (IPM).

\bibliographystyle{unsrt}

\bibliography{references}

\begin{thebibliography}{10}

\bibitem{world2019world}
{WHO}.
\newblock {\em {World Malaria Report 2019}}.
\newblock World Health Organization, Geneva, 2019.

\bibitem{barmania2015madagascar}
Sima Barmania.
\newblock Madagascar's health challenges.
\newblock {\em The Lancet}, 386(9995):729--730, 2015.

\bibitem{bhatt2015effect}
Samir Bhatt, DJ~Weiss, E~Cameron, D~Bisanzio, B~Mappin, U~Dalrymple, KE~Battle,
  CL~Moyes, A~Henry, PA~Eckhoff, et~al.
\newblock The effect of malaria control on \emph{{Plasmodium} falciparum} in
  africa between 2000 and 2015.
\newblock {\em Nature}, 526(7572):207--211, 2015.

\bibitem{howes2016contemporary}
Rosalind~E Howes, Sedera~Aur{\'e}lien Mioramalala, Brune Ramiranirina, Thierry
  Franchard, Andry~Joeliarijaona Rakotorahalahy, Donal Bisanzio, Peter~W
  Gething, Peter~A Zimmerman, and Ars{\`e}ne Ratsimbasoa.
\newblock Contemporary epidemiological overview of malaria in {Madagascar}:
  operational utility of reported routine case data for malaria control
  planning.
\newblock {\em Malaria Journal}, 15(1):502, 2016.

\bibitem{kang2018spatio}
Su~Yun Kang, Katherine~E Battle, Harry~S Gibson, Ars{\`e}ne Ratsimbasoa,
  Milijaona Randrianarivelojosia, St{\'e}phanie Ramboarina, Peter~A Zimmerman,
  Daniel~J Weiss, Ewan Cameron, Peter~W Gething, and Rosalind~E Howes.
\newblock Spatio-temporal mapping of {Madagascar’s Malaria Indicator Survey
  results to assess \emph{{Plasmodium} falciparum}} endemicity trends between
  2011 and 2016.
\newblock {\em BMC Medicine}, 16(1):71, 2018.

\bibitem{ihantamalala2018spatial}
Felana~A Ihantamalala, Feno~MJ Rakotoarimanana, Tanjona Ramiadantsoa,
  Jean~Marius Rakotondramanga, Gwena{\"e}lle Pennober, Fanjasoa Rakotomanana,
  Simon Cauchemez, Charlotte~JE Metcalf, Vincent Herbreteau, and Amy
  Wesolowski.
\newblock Spatial and temporal dynamics of malaria in madagascar.
\newblock {\em Malaria Journal}, 17(1):58, 2018.

\bibitem{randrianasolo2010sentinel}
Laurence Randrianasolo, Yolande Raoelina, Maherisoa Ratsitorahina, Lisette
  Ravolomanana, Soa Andriamandimby, Jean-Michel Heraud, Fanjasoa Rakotomanana,
  Robinson Ramanjato, Armand~Eug{\`e}ne Randrianarivo-Solofoniaina, and Vincent
  Richard.
\newblock Sentinel surveillance system for early outbreak detection in
  {Madagascar}.
\newblock {\em BMC Public Health}, 10(1):31, 2010.

\bibitem{nguyen2020mapping}
Michele Nguyen, Rosalind~E Howes, Tim~CD Lucas, Katherine~E Battle, Ewan
  Cameron, Harry~S Gibson, Jennifer Rozier, Suzanne Keddie, Emma Collins, Rohan
  Arambepola, et~al.
\newblock Mapping malaria seasonality in madagascar using health facility data.
\newblock {\em BMC Medicine}, 18(1):1--11, 2020.

\bibitem{nsp18}
{National Malaria Control Programme of Madagascar}.
\newblock Plan strategique national de lutte contre le paludisme: Elimination
  progressive du paludisme \`a madagascar, 2018.

\bibitem{president2019mdg}
{President's Malaria Initiative}.
\newblock Madagascar malaria operational plan financial year 2019, 2018.

\bibitem{president2018mdg}
{President's Malaria Initiative}.
\newblock Madagascar malaria operational plan financial year 2018, 2017.

\bibitem{battle2016treatment}
Katherine~E Battle, Donal Bisanzio, Harry~S Gibson, Samir Bhatt, Ewan Cameron,
  Daniel~J Weiss, Bonnie Mappin, Ursula Dalrymple, Rosalind~E Howes, Simon~I
  Hay, and Peter~W Gething.
\newblock Treatment-seeking rates in malaria endemic countries.
\newblock {\em Malaria Journal}, 15(1):20, 2016.

\bibitem{mis2013}
{Institut National de la Statistique (INSTAT), Programme National de lutte
  contre le Paludisme (PNLP), Institut Pasteur de Madagascar (IPM), and ICF
  International}.
\newblock {\em Madagascar Malaria Indicator Survey 2013 [Enquête sur les
  Indicateurs du Paludisme (EIPM)]}.
\newblock INSTAT, PNLP, IPM and ICF International, Calverton, 2013.

\bibitem{mis2016}
{Institut National de la Statistique (INSTAT), Programme National de lutte
  contre le Paludisme (PNLP), Institut Pasteur de Madagascar (IPM), and ICF
  International}.
\newblock {\em Madagascar Malaria Indicator Survey 2016 [Enquête sur les
  Indicateurs du Paludisme (EIPM)]}.
\newblock INSTAT, PNLP, IPM and ICF International, Calverton, 2016.

\bibitem{bennett2014methodological}
Adam Bennett, Joshua Yukich, John~M Miller, Penelope Vounatsou, Busiku
  Hamainza, Mercy~M Ingwe, Hawela~B Moonga, Mulakwo Kamuliwo, Joseph Keating,
  Thomas~A Smith, Richard~W Steketee, and Thomas~P Eisele.
\newblock A methodological framework for the improved use of routine health
  system data to evaluate national malaria control programs: evidence from
  zambia.
\newblock {\em Population health metrics}, 12(1):30, 2014.

\bibitem{chanda2012impact}
Emmanuel Chanda, Michael Coleman, Immo Kleinschmidt, Janet Hemingway, Busiku
  Hamainza, Freddie Masaninga, Pascalina Chanda-Kapata, Kumar~S Baboo, David~N
  D{\"u}rrheim, and Marlize Coleman.
\newblock Impact assessment of malaria vector control using routine
  surveillance data in zambia: implications for monitoring and evaluation.
\newblock {\em Malaria Journal}, 11(1):437, 2012.

\bibitem{mis2011}
{Institut National de la Statistique (INSTAT), Programme National de lutte
  contre le Paludisme (PNLP), and ICF International}.
\newblock {\em Madagascar Malaria Indicator Survey 2011 [Enquête sur les
  Indicateurs du Paludisme (EIPM)]}.
\newblock INSTAT, PNLP, IPM and ICF International, Calverton, 2012.

\bibitem{battle2019mapping}
Katherine~E Battle, Tim~CD Lucas, Michele Nguyen, Rosalind~E Howes, Anita~K
  Nandi, Katherine~A Twohig, Daniel~A Pfeffer, Ewan Cameron, Puja~C Rao, Daniel
  Casey, et~al.
\newblock Mapping the global endemicity and clinical burden of
  \emph{{Plasmodium vivax}}, 2000--17: a spatial and temporal modelling study.
\newblock {\em The Lancet}, 394(10195):332--343, 2019.

\bibitem{weiss2019mapping}
Daniel~J Weiss, Tim~CD Lucas, Michele Nguyen, Anita~K Nandi, Donal Bisanzio,
  Katherine~E Battle, Ewan Cameron, Katherine~A Twohig, Daniel~A Pfeffer,
  Jennifer~A Rozier, et~al.
\newblock Mapping the global prevalence, incidence, and mortality of
  \emph{{Plasmodium falciparum}}, 2000--17: a spatial and temporal modelling
  study.
\newblock {\em The Lancet}, 394(10195):322--331, 2019.

\bibitem{sturrock2016mapping}
Hugh~JW Sturrock, Adam~F Bennett, Alemayehu Midekisa, Roly~D Gosling, Peter~W
  Gething, and Bryan Greenhouse.
\newblock Mapping malaria risk in low transmission settings: challenges and
  opportunities.
\newblock {\em Trends in parasitology}, 32(8):635--645, 2016.

\bibitem{lucas2019model}
Tim~CD Lucas, Anita Nandi, Michele Nguyen, Susan Rumisha, Katherine~E Battle,
  Rosalind~E Howes, Chantal Hendriks, Andre Python, Penny Hancock, Ewan
  Cameron, et~al.
\newblock Model ensembles with different response variables for base and meta
  models: malaria disaggregation regression combining prevalence and incidence
  data.
\newblock {\em BioRxiv}, page 548719, 2019.

\bibitem{weiss2014air}
Daniel~J Weiss, Samir Bhatt, Bonnie Mappin, Thomas~P Van~Boeckel, David~L
  Smith, Simon~I Hay, and Peter~W Gething.
\newblock Air temperature suitability for \emph{{Plasmodium} falciparum}
  malaria transmission in {Africa} 2000-2012: a high-resolution spatiotemporal
  prediction.
\newblock {\em Malaria Journal}, 13(1):171, 2014.

\bibitem{weiss2018global}
D~J Weiss, A~Nelson, HS~Gibson, W~Temperley, S~Peedell, A~Lieber, M~Hancher,
  E~Poyart, S~Belchior, N~Fullman, B~Mappin, U~Dalrymple, J~Rozier, T~C~D
  Lucas, R~E Howes, L~S Tusting, S~Y Kang, E~Cameron, D~Bisanzio, K~E Battle,
  S~Bhatt, and P~W Gething.
\newblock A global map of travel time to cities to assess inequalities in
  accessibility in 2015.
\newblock {\em Nature}, 553(7688):333--336, 2018.

\bibitem{lucas2020mapping}
Tim~CD Lucas, Anita~K Nandi, Elisabeth~G Chestnutt, Katherine~A Twohig,
  Suzanne~H Keddie, Emma~L Collins, Rosalind~E Howes, Michele Nguyen, Susan
  Rumisha, Andre Python, Rohan Arambepola, Amelia Bertozzi-Villa, Penelope
  Hancock, Punam Amratia, Katherine~E Battle, Ewan Cameron, Peter~W Gething,
  and Daniel~J Weiss.
\newblock Mapping malaria by sharing spatial information between incidence and
  prevalence datasets.
\newblock {\em medRxiv}, 2020.

\bibitem{cameron2015defining}
Ewan Cameron, Katherine~E Battle, Samir Bhatt, Daniel~J Weiss, Donal Bisanzio,
  Bonnie Mappin, Ursula Dalrymple, Simon~I Hay, David~L Smith, Jamie~T Griffin,
  Edward~A Wenger, Philip~A Eckhoff, Thomas~A Smith, Melissa~A Penny, and
  Peter~W Gething.
\newblock Defining the relationship between infection prevalence and clinical
  incidence of \emph{{Plasmodium} falciparum} malaria.
\newblock {\em Nature communications}, 6(1):1--10, 2015.

\bibitem{liu2009modularization}
Fei Liu, MJ~Bayarri, and JO~Berger.
\newblock Modularization in bayesian analysis, with emphasis on analysis of
  computer models.
\newblock {\em Bayesian Analysis}, 4(1):119--150, 2009.

\bibitem{jacob2017better}
Pierre~E Jacob, Lawrence~M Murray, Chris~C Holmes, and Christian~P Robert.
\newblock Better together? statistical learning in models made of modules.
\newblock {\em arXiv preprint arXiv:1708.08719}, 2017.

\bibitem{elvidge2017viirs}
Christopher~D Elvidge, Kimberly Baugh, Mikhail Zhizhin, Feng~Chi Hsu, and
  Tilottama Ghosh.
\newblock Viirs night-time lights.
\newblock {\em International Journal of Remote Sensing}, 38(21):5860--5879,
  2017.

\bibitem{arambepola2020nonparametric}
Rohan Arambepola, Peter Gething, and Ewan Cameron.
\newblock Nonparametric causal feature selection for spatiotemporal risk
  mapping of malaria incidence in {Madagascar}.
\newblock {\em arXiv preprint arXiv:2001.07745}, 2020.

\bibitem{funk2014quasi}
Chris~C Funk, Pete~J Peterson, Martin~F Landsfeld, Diego~H Pedreros, James~P
  Verdin, James~D Rowland, Bo~E Romero, Gregory~J Husak, Joel~C Michaelsen, and
  Andrew~P Verdin.
\newblock A quasi-global precipitation time series for drought monitoring.
\newblock {\em US Geological Survey Data Series}, 832(4):1--12, 2014.

\bibitem{modisLST}
{NASA Earth Observations}.
\newblock Average land surface temperature.
\newblock
  \url{http://neo.sci.gsfc.nasa.gov/view.php?datasetId=MOD\_LSTD\_CLIM\_M},
  2017.
\newblock [Accessed Sept 2017].

\bibitem{modisTCB}
{NASA Earth Data}.
\newblock Land processes distributed active archive center.
\newblock
  \url{https://lpdaac.usgs.gov/dataset_discovery/modis/modis_products_table/mcd43b5},
  2017.
\newblock [Accessed Sept 2017].

\bibitem{modisEVI}
{NASA Earth Data}.
\newblock {MODIS (MOD 13) - Gridded} vegetation indices {(NDVI and EVI)}.
\newblock
  \url{http://modis.gsfc.nasa.gov/data/dataprod/dataproducts.php?MOD\_NUMBER=13},
  2017.
\newblock [Accessed Sept 2017].

\bibitem{trabucco2009global}
Antonio Trabucco and Robert~J Zomer.
\newblock Global aridity index (global-aridity) and global potential
  evapo-transpiration (global-pet) geospatial database.
\newblock {\em CGIAR Consortium for Spatial Information}, 2009.

\bibitem{farr2007shuttle}
Tom~G Farr, Paul~A Rosen, Edward Caro, Robert Crippen, Riley Duren, Scott
  Hensley, Michael Kobrick, Mimi Paller, Ernesto Rodriguez, Ladislav Roth,
  David Seal, Scott Shaffer, Joanne Shimada, Jeffrey Umland, Marian Werner,
  Michael Oskin, Douglas Burbank, and Douglas Alsdorf.
\newblock The shuttle radar topography mission.
\newblock {\em Reviews of geophysics}, 45(2), 2007.

\bibitem{lehner2004global}
Bernhard Lehner and Petra D{\"o}ll.
\newblock Global lakes and wetlands database glwd.
\newblock {\em GLWD Documentation}, 2004.

\bibitem{dijkstra1959note}
Edsger~W Dijkstra.
\newblock A note on two problems in connexion with graphs.
\newblock {\em Numerische mathematik}, 1(1):269--271, 1959.

\bibitem{pach2016qualitative}
Alfred Pach, Michelle Warren, Irene Chang, Justin Im, Chelsea Nichols,
  Christian~G Meyer, Gi~Deok Pak, Ursula Panzner, Se~Eun Park, Vera von
  Kalckreuth, Stephen Baker, Henintsoa Rabezanahary, Jean~Philibert
  Rakotondrainiarivelo, Tiana~Mirana Raminosoa, Rakotozandrindrainy Rapha\"el,
  and Florian Marks.
\newblock A qualitative study investigating experiences, perceptions, and
  healthcare system performance in relation to the surveillance of typhoid
  fever in madagascar.
\newblock {\em Clinical Infectious Diseases}, 62(suppl\_1):S69--S75, 2016.

\bibitem{alegana2012spatial}
Victor~A Alegana, Jim~A Wright, Uusiku Pentrina, Abdisalan~M Noor, Robert~W
  Snow, and Peter~M Atkinson.
\newblock Spatial modelling of healthcare utilisation for treatment of fever in
  namibia.
\newblock {\em International journal of health geographics}, 11(1):6, 2012.

\bibitem{GPWv4}
{NASA} Socioeconomic~Data Center~for International Earth Science Information
  Network CIESIN Columbia~University and Applications~Center (SEDAC).
\newblock Gridded population of the world, version 4 (gpwv4): {A}dministrative
  unit center points with population estimates, 2018.

\bibitem{tatem2017worldpop}
Andrew~J Tatem.
\newblock Worldpop, open data for spatial demography.
\newblock {\em Scientific data}, 4(1):1--4, 2017.

\bibitem{moh2014}
{Ministry of Health of Madagascar}.
\newblock Reference manual of principal health system indicators (in french).
  antananarivo: Ministry of health of madagascar, 2014.

\bibitem{lindgren2011explicit}
Finn Lindgren, H{\aa}vard Rue, and Johan Lindstr{\"o}m.
\newblock An explicit link between {Gaussian fields and Gaussian Markov} random
  fields: the stochastic partial differential equation approach.
\newblock {\em Journal of the Royal Statistical Society: Series B (Statistical
  Methodology)}, 73(4):423--498, 2011.

\bibitem{kristensen2015template}
K~Kristensen, A~Nielsen, CW~Berg, H~Skaug, and B~Bell.
\newblock Template model builder {TMB}.
\newblock {\em J. Stat. Softw}, 70:1--21, 2015.

\bibitem{guyon2007causal}
Isabelle Guyon and Constantin Aliferis.
\newblock Causal feature selection.
\newblock In {\em Computational methods of feature selection}, pages 79--102.
  Chapman and Hall/CRC, 2007.

\bibitem{spirtes2000causation}
Peter Spirtes, Clark~N Glymour, Richard Scheines, and David Heckerman.
\newblock {\em Causation, prediction, and search}.
\newblock MIT press, 2000.

\bibitem{strobl2019approximate}
Eric~V Strobl, Kun Zhang, and Shyam Visweswaran.
\newblock Approximate kernel-based conditional independence tests for fast
  non-parametric causal discovery.
\newblock {\em Journal of Causal Inference}, 7(1), 2019.

\bibitem{scholkopf2012causal}
Bernhard Sch{\"o}lkopf, Dominik Janzing, Jonas Peters, Eleni Sgouritsa, Kun
  Zhang, and Joris Mooij.
\newblock On causal and anticausal learning.
\newblock {\em arXiv preprint arXiv:1206.6471}, 2012.

\bibitem{president2017mdg}
{President's Malaria Initiative}.
\newblock Madagascar malaria operational plan financial year 2017, 2016.

\bibitem{president2013mdg}
{President's Malaria Initiative}.
\newblock Madagascar malaria operational plan financial year 2013, 2012.

\bibitem{president2014mdg}
{President's Malaria Initiative}.
\newblock Madagascar malaria operational plan financial year 2014, 2013.

\bibitem{do2018associations}
Mai Do, Stella Babalola, Grace Awantang, Michael Toso, Nan Lewicky, and Andrew
  Tompsett.
\newblock Associations between malaria-related ideational factors and
  care-seeking behavior for fever among children under five in mali, nigeria,
  and madagascar.
\newblock {\em PloS one}, 13(1), 2018.

\bibitem{alegana2020routine}
Victor~A Alegana, Emelda~A Okiro, and Robert~W Snow.
\newblock Routine data for malaria morbidity estimation in africa: challenges
  and prospects.
\newblock {\em BMC Medicine}, 18(1):1--13, 2020.

\end{thebibliography}

\end{document}